\documentclass{article}

\usepackage[nonatbib,preprint]{neurips_2019}
\usepackage[utf8]{inputenc} 
\usepackage[T1]{fontenc}    
\usepackage{hyperref}       
\usepackage{url}            
\usepackage{booktabs}       
\usepackage{amsfonts}       
\usepackage{nicefrac}       
\usepackage{microtype}      
\usepackage[numbers]{natbib}

\usepackage{amsmath}
\usepackage{mathtools}
\usepackage{amsthm}
\usepackage{amssymb}
\usepackage{wasysym}
\usepackage{graphicx}
\usepackage{subfig}
\title{Signal Coding and Perfect Reconstruction using Spike Trains}

\author{%
Anik Chattopadhyay\\ 
  Computer \& Information Science \& Engineering\\
  University of Florida\\
  Gainesville, FL 32611 \\
  \texttt{achattopadhyay@ufl.edu} \\
   \And
  Arunava Banerjee \\
  Computer \& Information Science \& Engineering\\
  University of Florida\\
  Gainesville, FL 32611 \\
  \texttt{arunava@ufl.edu} \\
}

\begin{document}

\maketitle
\begin{abstract}
In many animal sensory pathways, the transformation from external stimuli to spike trains is essentially deterministic. In this context, a new mathematical framework for coding and reconstruction, based on a biologically plausible model of the spiking neuron, is presented. The framework considers encoding of a signal through spike trains generated by an ensemble of neurons via a standard convolve-then-threshold mechanism. Neurons are distinguished by their convolution kernels and threshold values. Reconstruction is posited as a convex optimization minimizing energy. Formal conditions under which perfect reconstruction of the signal from the spike trains is possible are then identified in this setup. Finally, a stochastic gradient descent mechanism is proposed to achieve these conditions. Simulation experiments are presented to demonstrate the strength and efficacy of the framework.
\end{abstract}

\section{Introduction}

Spike based encoding of sensory stimuli is a hallmark of biological systems. It is now well-established that the coding of continuous time sensory signals in spike trains is a complex and diverse phenomenon, and is fairly deterministic in many animal sensory pathways\citep{meister1995, christopher1996, neuenschwander1996, wehr1996, johansson2004, nemenman2008}. Spike train representations, when sparse, are not only intrinsically energy efficient, but can also facilitate computation at later stages of processing\citep{Foldiak1990, Graham}. In their seminal work, Olshausen and Field \citep{Olshausen1996} showed how efficient codes can arise from learning sparse representations of natural stimulus statistics, resulting in striking similarities with observed biological receptive fields. Smith and Lewicki \citep{Lewicki2002, CSmith2006} likewise showed that auditory filters could be estimated by training a population spike code model with natural sounds. These studies, by and large, fall under the general framework of dictionary learning: identifying an over-complete dictionary $\{\phi_j|j=1\ldots m\}$ such that each stimulus $s_i$ in an ensemble $\{s_i|i=1\ldots n\}$ can be represented as $s_i=\sum_{j=1}^m \alpha_j \phi_j$ where the vector of coefficients $\alpha_j$ is sparse. The studies only made passing reference to how the $\alpha_j$'s may be derived (e.g. matching pursuit \cite{Mallat1993}) or even be represented (e.g. local population of neurons spiking probabilistically proportional to $\alpha_j$ in \cite{CSmith2006}). Lacking clearly specified plausible neural implementations, the extent to which the proposed schemes underlie biological sensory processing therefore remained unclear. To remedy this, several subsequent learning techniques based on biologically plausible models of spiking neurons have been proposed. For example, \citep{murphy} developed a biophysically motivated spiking neural network which for the first time predicted the full diversity of V1 simple cell receptive field shapes when trained on natural images. Elsewhere \citep{shapero} presented a rate encoded spiking neural network of integrate-and-fire neurons demonstrating convergence to nearly optimal encodings.

Although these results signify substantial progress, the classical signal processing question of what class of signals support perfect or approximate reconstruction when coded using spike trains, remains to be fully resolved. Admittedly, the very coarse $\Sigma\Delta$ quantization of bandlimited signals investigated in \citep{daubechies} does amount to a spike train representation. However, due to the classical nature of its framework, not only is biological plausibility not a concern, but also coding is explored in the oversampled regime. Along similar lines, \citep{chklovskii} has explored the spike generating mechanism of the neuron as an oversampling, noise shaping analog-to-digital converter.

Here we present a new framework for coding and reconstruction that begins with a biologically plausible coding mechanism which is a superset of the standard leaky integrate-and-fire mechanism. Reconstruction is first formulated as an optimization that minimizes the energy of the reconstructed signal subject to consistency with the spike train, and then solved in closed form. We then identify a general class of signals for which reconstruction is provably perfect under certain conditions. Surprisingly, the result instantiates a version of Barlow's ``efficient coding hypothesis'' \citep{barlow1961}, which posits that the coding strategy of sensory neurons should be adapted to the statistics of the stimuli in an animal's natural environment. We present a stochastic gradient descent mechanism to achieve these conditions, and close with simulation experiments that demonstrate the efficacy of the framework. The rest of the paper is structured as follows. In Sections~\ref{Encoding Module} and \ref{Decoding Module} we introduce the coding and decoding frameworks. Section~\ref{Class} identifies the class of signals for which perfect reconstruction is achievable if certain conditions are met. A learning scheme to achieve these conditions is presented in Section~\ref{Learning}, followed by simulation results in Section~\ref{Experiments}. We conclude in Section~\ref{Conclusion}.

\section{Coding}
\label{Encoding Module}

The general class of deterministic transformations (i.e., the set of all nonlinear operators) from continuous time signals to spike trains is difficult to characterize because the space of all spike trains does not lend itself to a natural topology that is universally agreed upon. The result is that simple characterizations, such as the set of all continuous operators, can not be posited in a manner that has general consensus. To resolve this issue, we take a que from biological systems. In most animal sensory pathways, external stimulus passes through a series of transformations before being turned into spike trains\citep{squire}. For example, visual signal in the retina is processed by multiple layers of non-spiking horizontal, amacrine and bipolar cells, before being converted into spike trains by the retinal ganglion cells. Accordingly, we can consider the set of transformations that pass via an intermediate continuous time signal which is then transformed into a spike train through a simple stereotyped mapping where spikes mark threshold crossings. The complexity of the operator now lies in the mapping from the continuous time input signal to the continuous time intermediate signal. Since any time invariant, continuous, nonlinear operator with fading memory can be approximated by a finite Volterra series operator\citep{boyd1985}, this general class of nonlinear operators from continuous time signals to spike trains can be modeled as the composition of a finite Volterra series operator and a neuronal thresholding operation to generate a spike train.  In our encoding model the simplest subclass of these transformations is considered: the case where the Volterra series operator has a single causal, bounded-time, {\em linear} term, the output of which is composed with a thresholding operation of a potentially time varying threshold. The overall operator from the input signal to the spike train remains {\em nonlinear} due to the thresholding operation.

Formally, we assume the input signal $X(t)$ to be a \emph{bounded continuous function} in the interval $[0,L]$ for some $L \in R^{+}$, i.e., we are interested in the class of input signals $\mathcal{F} = \{X(t)| X(t) \in C[0,L]\}$. Since the framework involves signal snippets of arbitrary length, this choice of $L$ is without loss of generalization. We assume an ensemble of convolution kernels $K= \{ K^j| j\in Z^+, j \leq n \} $, consisting of $n$ kernels $K^j, j=1,\ldots,n$. We assume that $K^j(t)$ is a continuous function on a bounded time interval $[0,T]$, i.e. $\forall j \in \{1,\ldots,n\}, K^j(t) \in C[0,T]$ for some $T \in R^+$. Finally, we assume that $K^j$ has a time varying threshold denoted by $T^j(t)$.  

The ensemble of convolution kernels $K$ encodes a given input signal $X(t)$ into a sequence of spikes $\{(t_i, K^{j_i})\}$, where the $i^{th}$ spike is produced by the $j_i^{th}$ kernel $K^{j_i}$ at time $t_i$ if and only if:
\begin{equation}
\label{spikeConstraint}
	\int X(\tau) K^{j_i}(t_i-\tau) d\tau = T^{j_i}(t_i) 
\end{equation}
We assume that the time varying threshold $T^j(t)$ of the $j{th}$ kernel remains constant at $C^j$ until that kernel produces a spike, at which time an \emph{after-hyperpolarization potential (ahp)} kicks in to raise the threshold to a high value $M^j \gg C^{j}$, which then drops back linearly to its original value within a refractory period $\delta_{j}$. Formally, the threshold function $T^j(t)$ of the $j^{th}$ kernel is given by:
\begin{equation}
\label{thresholdeq}
T^j(t) = 
    \begin{cases*}
        C^j, &  $t-\delta_{j} > t_l^j(t)$ \\ 
        M^j - \frac{(t-t_l^j(t))(M^j-C^j)}{\delta_{j}}, & $t-\delta_{j} \leq  t_l^j(t) $\\
    \end{cases*}
\end{equation}
\hspace{50pt}\text{ Where  $t_l^j(t)$ denotes the time of the last spike generated by $K^{j}$ prior to time $t$}. \\
Notably, apart from the contribution due to the \emph{ahp}, we have considered the threshold of the $j^{th}$ kernel (alternately called a neuron) to be a constant $C^j$ in our model. This is in contrast to real neurons that are known to vary their thresholds through homeostasis.

\section{Decoding}
\label{Decoding Module}

The objective of the decoding module is to reconstruct the original signal from the encoded spike trains. Considering the prospect of the invertibility of the coding scheme, we seek a signal that satisfies the same set of constraints as the original signal when generating all spikes apropos the set of kernels in ensemble $K$. Recognizing that such a signal might not be unique, we choose the reconstructed signal as the one with minimum $L2$-norm. Formally, the reconstruction (denoted by $X^*(t)$) of the input signal $X(t)$ is formulated to be the solution to the optimization problem:\\
\begin{equation}
\label{optimizationproblem}
\begin{aligned}
	& X^{*}(t)= \underset{\Tilde{X}}{\text{argmin}}
	||\Tilde{X}(t)||_2^2 \\
	& \text{s.t.}
	\int \Tilde{X}(\tau)K^{j_i}(t_i-\tau)d\tau = T^{j_i}(t_i); 
	1 \le i \le N
\end{aligned}
\end{equation}
\text {where  $\{(t_i, K^{j_i})|i \in \{1,...,N\}\}$ is the set of all spikes generated by the encoder.}
\paragraph{Why L2 minimization?} The choice of $L2$ minimization as the objective of the reconstruction problem can only be weakly justified at the current juncture. The perfect reconstruction theorem that follows provides the strong justification. As it stands, the $L2$ minimization objective is in congruence with the dictum of energy efficiency in biological systems. The assumption is that, of all signals, the one with the minimum energy that is consistent with the spike trains is desirable. Secondly, an $L2$ minimization in the objective of (\ref{optimizationproblem}) reduces the convex optimization problem to a solvable linear system of equations as shown in Lemma 1. Later we shall show that $L2$-minimization has the surprising benefit of recovering the original signal perfectly under certain conditions.

\section{Signal Class for Perfect Reconstruction}
\label{Class}

To establish the effectiveness of the described coding-decoding model, we have to evaluate the accuracy of reconstruction over a class of input signals. We observe that in general the encoding of continuous time signals into spike trains is not a one-to-one map; the same set of spikes can be generated by different continuous time signals with appropriate changes in amplitudes so as to result in the same convolved values at the spike times. Naturally, with a finite and fixed ensemble of kernels $K$, one cannot achieve perfect reconstruction for the general class of signals $\mathcal{F}$ as defined in Section~\ref{Encoding Module}. We now restrict ourselves to a subset $\mathcal{G}$ of the original class $\mathcal{F}$ as defined below and address the question of reconstruction accuracy. 
\begin{align}
\label{restricted class}
\mathcal{G} = \{X(t)| X(t) \in \mathcal{F}, X(t)= \sum_{p=1}^{N} \alpha_p K^{j_p}(t_p-t), j_p \in \{1,...,n\}, \alpha_p \in R, t_p \in R^{+}, N \in Z^{+}\}
\end{align}
Essentially $\mathcal{G}$ consists of all linear combinations of arbitrarily shifted inverted kernel functions. $N$ is bounded above by the total number of spikes that the ensemble $K$ can generate over $[0,L]$. For this restricted class of signals the \emph{perfect reconstruction theorem} is presented below. The theorem is proved with the help of two lemmas.
\paragraph{Perfect Reconstruction Theorem:}
\label{PerfectReconsThm}
Let $X(t) \in \mathcal{G}$ be an input signal. Then for appropriately chosen time-varying thresholds of the kernels, the reconstruction, $X^{*}(t)$, resulting from the proposed coding-decoding framework is accurate with respect to the $L2$ metric, i.e., $||X^{*}(t)-X(t)||_2 = 0$. 
\paragraph{Lemma1:} 
\label{th1}
The solution $X^*(t)$ to the reconstruction problem given by (\ref{optimizationproblem}) can be written as:\\
\begin{equation}
\label{reconstructionequation}
\begin{aligned}
X^*(t) = \sum_{i=1}^{N} \alpha_i K^{j_i}(t_i-t)\\
\end{aligned}
\end{equation}
where the coefficients $\alpha_i \in R$ can be uniquely solved from a system of linear equations if the shifted kernel functions $K^{j_i}(t_i-t)$ are \emph{linearly independent}.
\paragraph{Proof:}
Application of the Representer Theorem \citep{scholkopf2001} on (\ref{optimizationproblem}) directly results in:
\begin{equation}
\label{eq:reconsSignalEq}
\begin{aligned}
X^*(t) = \sum_{i=1}^{N} \alpha_i K^{j_i}(t_i-t)\\
\end{aligned}
\end{equation}
where the $\alpha_{i}$'s are real valued coefficients. In essence, the reconstructed signal $X^{*}(t)$ becomes a summation of the kernels, shifted to their respective times of generation of spikes, scaled by appropriate coefficients. 
Plugging (\ref{eq:reconsSignalEq}) into the constraints (\ref{optimizationproblem}) gives:
\begin{equation*}
	\forall_{1\le i \le N} ;
	\int \sum\limits_{k=1}^{N}\alpha_k K^{j_k}(t_k-t)
	K^{j_i}(t_i-t) d\tau = T^{j_i}(t_i) 
\end{equation*}
Setting $b_i = T^{j_i}(t_i)$ and
$P_{ik} = \int K^{j_k}(t_k-\tau) K^{j_i}(t_i-\tau) d\tau$ results in:
\begin{equation}
	\label{eq:finalReconsEq}
	\forall_{1 \le i \le N} ;
	\sum\limits_{k=1}^{N} P_{ik} \alpha_k = b_i
\end{equation}
Equation~(\ref{eq:finalReconsEq}) defines a system of $N$ equations in $N$ unknowns of the form:\\
\begin{equation}
\label{alphaeq}
P\alpha=T
\end{equation}
where
$\alpha = \langle\alpha_1, ..., \alpha_N\rangle^T$,
 $T = \langle T^{j_1}(t_1), ..., T^{j_N}(t_N)\rangle^T$ and $P$ is an
$N \times N$ matrix with elements
$P_{ik} = \int K^{j_k}(t_k-\tau) K^{j_i}(t_i-\tau) d\tau$. Clearly $P$ is the Gramian Matrix of the shifted kernels $\{K^{j_i}(t_i-t)|i \in 1,2,...,N \}$ in the Hilbert space with the standard inner product. Hence $\alpha$ has a unique solution if and only if $P$ is invertible. Finally, the Gram matrix $P$ is invertible if and only if the set of vectors $\{K^{j_i}(t_i-t)|i \in 1,2,...,N \}$ in Hilbert space are linearly independent. $\hfill\Box$
\paragraph{Import:}
The goal of the optimization problem is to find the best object in the feasible set. However, the application of the Representer Theorem, as above, converts the constraints into a determined system of unknowns and equations, turning the feasible set into a single point, effectively changing the optimization problem into a solvable system that results in a closed form solution for the $\alpha_i$'s. This implies that instead of solving (\ref{optimizationproblem}), we can solve for the reconstruction from $X^*(t) = \sum_{i=1}^{N} \alpha_i K^{j_i}(t_i-t)$, where $\alpha_{i}$ is the i-th element of $ \alpha = P^{-1} T$.

\paragraph{Lemma2:}
\label{th2}
Let $X^*(t)$ be the reconstruction of an input signal $X(t)$ and $\{(t_i, K^{j_i})\}_{i=1}^{N}$ be the set of spikes generated. Then, for any arbitrary signal $\Tilde{X}(t)$ within the span of $\{K^{j_i}(t_i-t)|i \in \{1,2,...,N\}\}$, i.e., the set of shifted inverted kernels at respective spike times, given by \begin{equation}
\begin{aligned}
\Tilde{X}(t) = \sum_{i=1}^{N} a_i K^{j_i}(t_i-t)\\
\end{aligned}
\end{equation}
the following inequality holds:
\begin{equation}
    ||X(t)- X^*(t)|| \leq ||X(t)-\Tilde{X}(t)||
\end{equation}
\paragraph{Proof:}
\begin{align}
&||X(t)-\Tilde{X}(t)|| = ||\underbrace{X(t)-X^{*}(t)}_\text{A}+\underbrace{X^{*}(t)-\Tilde{X}(t)}_\text{B}|| \nonumber\\
&\text{First,\ \ } \langle A,K^{j_i}(t_i-t) \rangle = \langle X(t), K^{j_i}(t_i-t) \rangle - \langle X^{*}(t), K^{j_i}(t_i-t) \rangle, \forall i \in \{1,2,..,N\} \nonumber\\
&\hspace{95pt} = T^{j_i}(t_i) -T^{j_i}(t_i) = 0  \hspace{50pt}\text{ (Using the constraints in (\ref{optimizationproblem}) \& (\ref{spikeConstraint}))} \nonumber\\
&\text{Second,\ \ } \langle A,B \rangle = \langle A, \sum_{i=1}^{N} (\alpha_i - a_i) K^{j_i}(t_i-t) \rangle \hspace{20pt}\text{ (By Lemma 1 $X^*(t) = \sum_{i=1}^{N} \alpha_i K^{j_i}(t_i-t)$)} \nonumber\\
&\hspace{55pt}= \sum_{i=1}^{N} (\alpha_i - a_i)\langle A, K^{j_i}(t_i-t) \rangle = 0 \nonumber\\
&\implies ||X(t)-\Tilde{X}(t)||^{2} = ||A+B||^{2} = ||A||^{2}+ 2 \langle A,B \rangle +||B||^{2}\\ &\hspace{100pt} =||A||^{2} + ||B||^{2} \geq ||A||^{2} = ||X(t)-X^{*}(t)||^{2} \nonumber\\
&\implies ||X(t)-\Tilde{X}(t)|| \geq ||X(t)-X^{*}(t)|| \nonumber \hspace{210pt}\Box
\end{align}
\paragraph{Import:} The implication of the above lemma is quite remarkable. The objective defined in (\ref{optimizationproblem}) chooses a signal with minimum energy satisfying the constraints, deemed the reconstructed signal. However as the lemma demonstrates, this signal also has the minimum \emph{ error} with respect to the input signal in the span of the shifted kernels. This signifies that our choice of the objective in the decoding module not only draws from biologically motivated energy optimization principles, but also performs optimally in terms of reconstructing the original input signal within the span of the appropriately shifted spike generating kernels.

Exploring further, for a given input signal $X(t)$ if $S_1$ and $S_2$ are two sets of spike trains where $S_1 \subset S_2$ produced by two different kernel ensembles, the second a superset of the first, then Lemma 2 further implies that the reconstruction due to $S_2$ is at least as good as the reconstruction due to $S_1$ because the reconstruction due to $S_1$ is in the span of the shifted kernel functions of $S_2$ as  $S_1 \subset S_2$. This immediately leads to the conclusion that for a given input signal the more kernels we add to the ensemble the better the reconstruction, provided the kernels maintain linear independence. 
\paragraph{Proof of the Theorem:}
The proof of the theorem follows directly from Lemma 2. Since the input signal $X(t) \in \mathcal{G}$, let $X(t)$ be given by the equation below:
\begin{align}
    X(t)= \sum_{p=1}^{N} \alpha_p K^{j_p}(t_p-t) \hspace{10pt}(\alpha_p \in R, t_p \in R^{+}, N \in Z^{+})
\end{align}
Assume that the time varying thresholds of the kernels in our kernel ensemble $K$ is set in such a manner that the below conditions are satisfied:
\begin{align}
    \langle X(t), K^{j_p}(t_p-t) \rangle = T^{j_p}(t_p) \hspace{10pt} \forall{p \in \{1,...,N\}}
\end{align}
i.e., each of the kernels $K^{j_p}$ at the very least produces a spike at time $t_p$ against $X(t)$ (regardless of other spikes at other times). Clearly then $X(t)$ lies in the span of the appropriately shifted and inverted response functions of the spike generating kernels. Applying Lemma 2 it follows that:
\begin{align}
    ||X(t)- X^*(t)||_2 \leq ||X(t)-X(t)||_2 = 0 \nonumber \hspace{210pt}\Box
\end{align}
\paragraph{Import:} In addition to demonstrating the potency of the coding-decoding scheme, this theorem frames Barlow's efficient coding hypothesis \citep{barlow1961}---that the coding strategy of sensory neurons be adapted to the statistics of the stimuli---in mathematically concrete terms. Going by the theorem, the spike based encoding necessitates the signals to be in the span of the encoding kernels for perfect reconstruction. Inverting the argument, kernels must learn to adapt to the \emph{basis} elements that generate the signal corpora for superior reconstruction. A practical challenge on which the reconstruction accuracy depends, as indicated by the theorem, is whether we can generate spikes at the correct temporal locations. One way to tackle this problem, as adopted in our experiments in Section~\ref{Experiments}, is to set the initial threshold values $C^{j}$s and the refractory periods $\delta_{j}$s to be low in our simple thresholding model (\ref{thresholdeq}). This then ensures that the real spikes deviate from their desired locations by at most $\delta_{j}$ time. Since the decoding is a continuous transformation, this changes the reconstruction accuracy by a small amount as is confirmed by our experiments in Section~\ref{Experiments}.
\section{Approximate Reconstruction of a Signal:}
\label{ImperfectRecons}
\begin{figure}[h]
 	\centering
 	\subfloat[]{{\includegraphics[width=50mm]{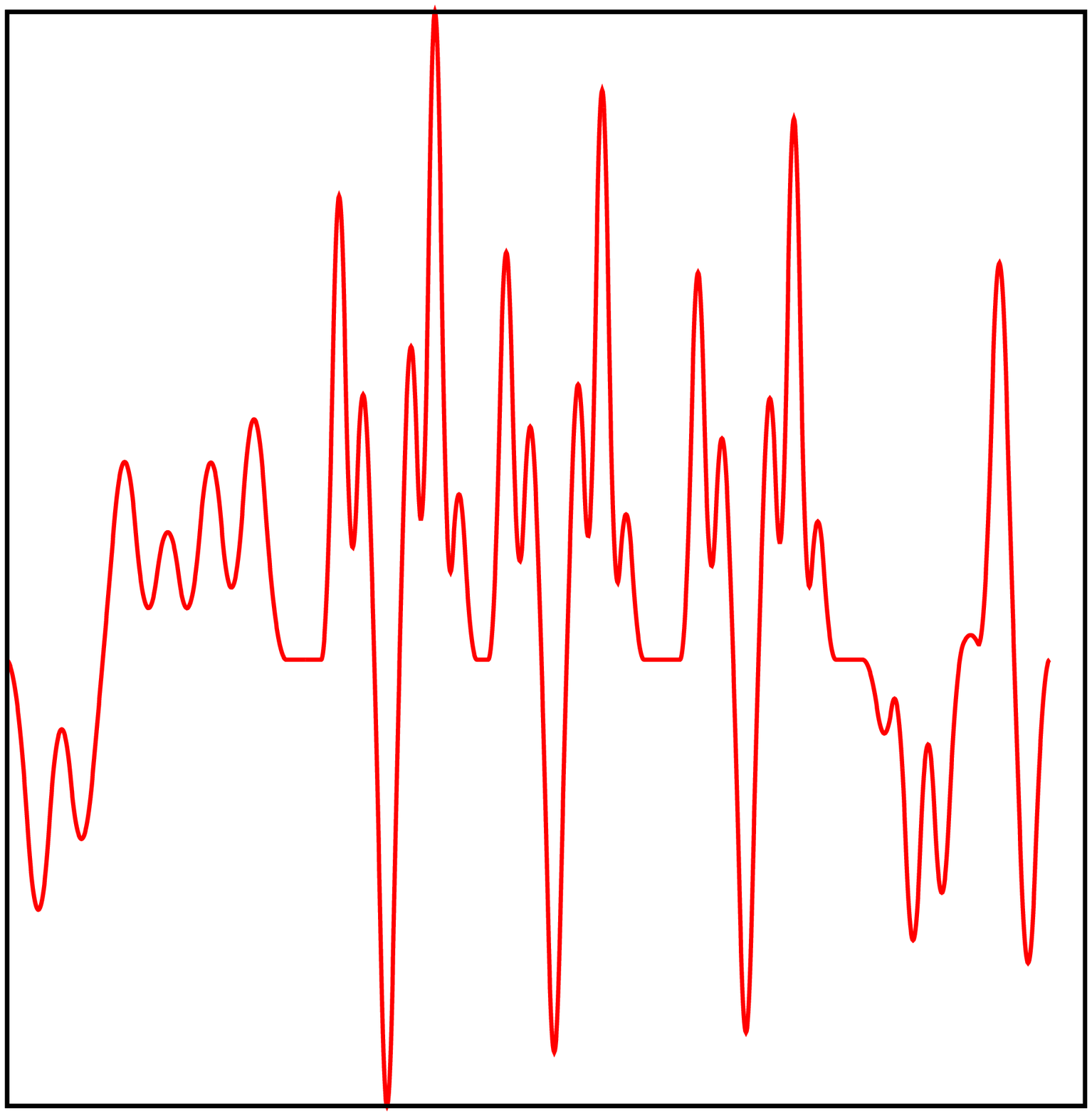} }}%
  	\subfloat[]{{\includegraphics[width=50mm]{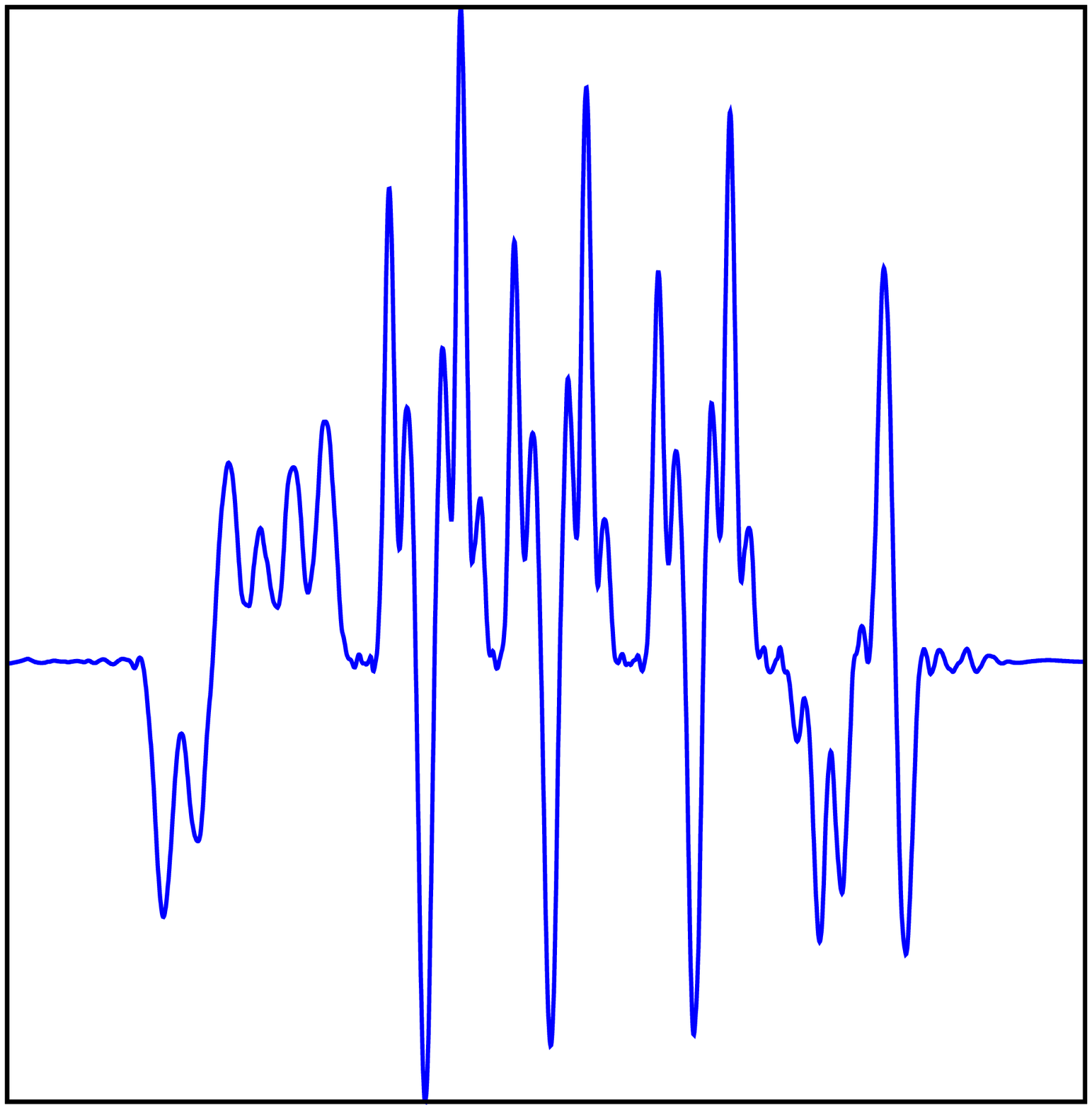}}}%
 	 \caption{ shows a sample reconstruction in an experiment with 10 kernels after training.
 	 (a) The original input signal is shown in red. (b) The reconstruction, shown in blue, is nearly perfect.}%
 	\label{PerfectReconstruction}
\end{figure}

The perfect reconstruction theorem \ref{PerfectReconstruction} essentially stipulates the exact conditions under which exact recovery of a signal is feasible in the proposed framework. But in real applications it could be challenging to meet those conditions for any arbitrary signal. For example, we may not be able to generate the spikes in exact times or the signal may not perfectly fit in the subspace generated from a finite bag of kernels, $\mathcal{G}$ as stated in the theorem \ref{PerfectReconstruction}. The goal of the approximate reconstruction theorem is to give a lower bound on the reconstruction error when the conditions are reasonably relaxed. 
\paragraph{Approximate Reconstruction Theorem:}
Let $X(t) \in L^{2}([0,L]), \text{for some } L \in R^{+}$, a square integrable continuous time signal in the interval $[0,L]$ be an input to our proposed framework. 
If the following assumptions are true:\\
\begin{itemize}
\item 
 $X(t)$ can be realized as a linear combination of some component signals as below:\\
\begin{equation}
    X(t) = \sum_{i=1}^{N} \alpha_i f_{p_i}(t-t_i)
\end{equation}
where $  p_{i}\in Z^{+}$, and components are chosen from a possible infinite set of functions of unit $L^{2}$ norm: $\{f_{i}(t)| i \in  Z^{+}, ||f_{i}(t)||_{2}= 1\}$ i.e. each component function $f_i(t)$ is normalized to have unit $L^{2}$ norm, and $\alpha_{i}$ are the real coefficients.\\
Also there $\exists$ a kernel $K^{j_i}$ from our bag of kernels K, such that $f_{p_i}$ is very close to $K^{j_i}(t)$ in $L^{2}$ norm i.e. $ \exists \delta \in R^{+} st ||f_{p_i}(t) - k^{j_i}(t)||_{2} < \delta \forall i \in \{ 1,...,N\}$.
\item 
Also assume that when $X(t)$ is encoded using our framework \ref{Encoding Module}, each one of these fitting kernels $k^{j_i}$ produces a spike at time $t_{i}'$ such that $|t_{i}- t_{i}'| < \Delta \forall i$. 
\item
Assume the response function of each kernel in our kernel bag satisfies a Lipschutz type of condition as follows, i.e. $\exists C \in R s.t. ||k^{j}(t)- k^{j}(t-\Delta t)||_{2} \leq C|\Delta t| \forall \Delta t \in R, \forall j$.
\item
Finally, we assume the shifted component functions also satisfy a frame bound type of condition as follows:
$
\sum_{k \neq i} <f_{p_i}(t-t_i),f_{p_k}(t-t_k)> \leq \eta \forall i \in {1,...,N}
$
\end{itemize}
Then, the noise to signal ratio in reconstruction of $X(t)$ resulting from the proposed coding-decoding framework is bounded. Specifically, the following inequality is satisfied:\\
\begin{equation}
\nicefrac{||X(t)-X_{hyp}(t)||_{2}^{2}}{||X(t)||_{2}^{2}} \leq (\delta  + C\Delta ) \nicefrac{(1+x_{max})}{(1-\eta)}
\end{equation}
\\where $x_{max}$ is a positive number $ \in [0, N-1]$ that depends on the maximum overlap of the support of component functions $f_{p_i}(t-t_i)$.  
\paragraph{Proof of the Theorem:}
\begin{align}
&\text{Let $X^{*}(t)$ be the reconstruction of input signal $X(t)$ using the proposed coding-decoding model. By} \nonumber\\
&\text{hypothesis each kernel $K^{j_i}$ produces a spike at time $t_{i}'$ which we have in some sense assumed to } \nonumber\\&\text{match the input signal. But our framework might generate some spurious spikes against $X(t)$. Other} \nonumber\\&\text{than the
set of spikes $\{(t_{i}', K^{j_i})|i \in \{1,...,N\}\}$, let
$\{(\Tilde{t_k}, K^{\Tilde{j_k}})|k \in \{1,...,M\}\}$
denote those extra } \nonumber\\
&\text{set of spikes that the coding-decoding model produces for input $X(t)$ against the kernel bag K. Here } \nonumber\\
&\text{we have assumed $M$ to be the number of spurious spikes.  
By lemma1 \ref{th1} $X^{*}(t)$ can be represented as}\nonumber\\
&\text{below:}\nonumber\\
&X^{*}(t) = 
\sum_{i=1}^{N} \alpha_{i} K^{j_i}(t -t_i') +\sum_{k=1}^{M} \Tilde{\alpha_k} K^{\Tilde{j_k}}(t -\Tilde{t_{k}}) \nonumber\\
&\text{where $\alpha_i$s and $\Tilde{\alpha_k}$s are real coefficients whose values can be formulated again from lemma1 (\ref{th1}).
Also } \nonumber\\
&\text{let $T_{i}$ be the threshold at which kernel  $K^{j_i}$ produced the spike at time $t_{i}'$ as given in the hypothesis.} \nonumber\\
&\text{Hence for generation of spike the below condition must be satisfied:} \nonumber\\ 
&<X(t), K^{j_i}(t-t_{i}')> = T_{i} \forall i \in \{1,2,...,N\} \nonumber\\
& \text{Consider a hypothetical signal $X_{hyp}(t)$ defined by the below equations:}\nonumber\\
& X_{hyp}(t) = \sum_{i=1}^{N} a_i K^{j_i}(t -t_{i}') , a_i \in R \nonumber\\
& s.t. <X_{hyp}(t), K^{j_i}(t - t_{i}')> = T_i \forall i \nonumber\\
&\text{By the formulation of lemma1 \ref{th1} it is obvious to see that such a signal is well defined because the} \nonumber\\
&\text{coefficients $a_{i}$s have unique solution.
Clearly the hypothetical signal can be deemed as if it is the} \nonumber \\
&\text{reconstructed signal where we are only considering the fitting spikes at times $t_{i}'$s and ignoring all}\nonumber\\
&\text{other spurious spikes that might be generated by our framework.
Since, $X_{hyp}(t)$ lies in the span of } \nonumber\\
&\text{shifted kernels used in reconstruction of
$X(t)$ using lemma2 \ref{th2} we may now write:} \nonumber\\
& ||X(t)-X_{hyp}(t)|| \geq ||X(t)-X^{*}(t)|| \label{part1}\\
&||X(t)-X_{hyp}(t)||_{2}^{2} = <X(t)-X_{hyp}(t), X(t)-X_{hyp}(t)> \nonumber \\
&= <X(t)-X_{hyp}(t), X(t)>-<X(t)-X_{hyp}(t), X_{hyp}(t)> \nonumber \\
& =<X(t)-X_{hyp}(t),X(t)> - \Sigma_{i=1}^{N} a_{i}<X(t)-X_{hyp}(t),K^{j_i}(t -t_i')> \nonumber \\ 
&= ||X(t)||_{2}^{2} - <X(t), X_{hyp}(t)> \nonumber \\
&(\because \text{by construction}  <X_{hyp}(t),K^{j_i}(t -t_i')>= <X(t),K^{j_i}(t -t_i')> = T_{i} \forall i\in{1,2,...,N}) \nonumber \\
&= \Sigma_{i=1}^{N}\Sigma_{k=1}^{N} \alpha_{i} \alpha_{k}<f_{i}(t-t_{i}),f_{k}(t-t_{k})>
- \Sigma_{i=1}^{N}\Sigma_{k=1}^{N} \alpha_{i} a_{k}<f_{i}(t-t_{i}),K^{j_k}(t -t_k'))> \nonumber\\
&= \alpha^{T}F\alpha - \alpha^{T}F_{K}a \label{error}\\
& (\text{denote } a= \begin{bmatrix} 
a_{1} \\
a_{2}\\
.\\
.\\
.\\
a_{N}\\
\end{bmatrix} , 
 \alpha = \begin{bmatrix} 
\alpha_{1}\\
\alpha_{2}\\
.\\
.\\
.\\
\alpha_{N}\\
\end{bmatrix} ) \nonumber\\
& (\text{denoting } F = [F_{ik}]_{NXN} \text{ where } F_{ik} = \langle f_{i}(t-t_{i}),f_{k}(t-t_{k}) \rangle \nonumber \\
&\text{ and } F_{K} = [(F_{K})_{ik}]_{NXN} \text{ where } (F_{K})_{ik} = \langle f_{i}(t-t_{i}), K^{j_k}(t-t_{k}')\rangle) \nonumber \\ 
&\text{But using the results of Lemma1 (\ref{th1}) $a$ can be written as: } \nonumber\\
&a = P^{-1}T \text{ where } P = [P_{ik}]_{NXN}, P_{ik} = <K^{j_i}(t-t_{i}'),K^{j_k}(t-t_{k}')> \nonumber\\
& \text{And, } T = [T_{i}]_{N\times 1} \text{ where } T_{i} = \langle X(t), K^{j_i}(t-t_{i}') \rangle = \Sigma_{k=1}^{N}\alpha_{k} \langle f_{k}(t-t_k), K^{j_i}(t-t_{i}') \rangle = F_{K}^{T}\alpha \nonumber\\
& \implies a = P^{-1}F_{K}^{T}\alpha \label{coeffs}\\
&\text{Combining equations \ref{error} and \ref{coeffs} we get,} \nonumber\\
&|X(t)-X_{hyp}(t)||_{2}^{2} = \alpha^{T}F\alpha - \alpha^{T}F_{K}P^{-1}F_{K}^{T}\alpha \label{expandedError}\nonumber\\
\end{align}
\begin{align}
& \text{But,} (F_{K})_{ik} = \langle f_{i}(t-t_i), K^{j_k}(t-t_k') \rangle \nonumber\\
&= \langle K^{j_i}(t-t_i'), K^{j_k}(t-t_k') \rangle - \langle K^{j_i}(t-t_i')-f_i(t-t_i), K^{j_k}(t-t_k') \rangle \nonumber\\
&= (P)_{ik} - (\mathcal{E}_{K})_{ik} \label{FKform}\\
&(\text{denoting } \mathcal{E}_{K} = [(\mathcal{E}_{K})_{ik}] \text{ where } (\mathcal{E}_{K})_{ik} = \langle K^{j_i}(t-t_i')-f_i(t-t_i), K^{j_k}(t-t_k') \rangle) \nonumber\\
&\text{Also,} \nonumber \\
&(F)_{ik} = \langle f_{i}(t-t_{i}), f_{k}(t-t_{k}) \rangle \nonumber\\
  & = \langle f_{i}(t-t_{i})-K^{j_i}(t-t_i')+K^{j_i}(t-t_i'), f_{k}(t-t_{k})-K^{j_k}(t-t_k')+K^{j_k}(t-t_k') \rangle \nonumber\\
&=  (\mathcal{E})_{ik} - (\mathcal{E}_{K})_{ik} - (\mathcal{E}_{K})_{ki} + (P)_{ik} \label{Fform}\\
& \text{Combining \ref{expandedError}, \ref{FKform} and \ref{Fform} we get,} \nonumber\\
&||X(t)-X_{hyp}(t)||_{2}^{2} = \alpha^{T}F\alpha - \alpha^{T}F_{K}P^{-1}F_{K}^{T}\alpha \nonumber\\
&= \alpha^{T} \mathcal{E} \alpha - \alpha^{T}\mathcal{E}_{K}\alpha - \alpha^{T}\mathcal{E}_{K}^{T}\alpha + \alpha^{T}P\alpha \nonumber\\
&- \alpha^{T}P\alpha + \alpha^{T}\mathcal{E}_{K}\alpha + \alpha^{T}\mathcal{E}_{K}^{T}\alpha - \alpha^{T}\mathcal{E}_{K}P{-1}\mathcal{E}_{K}^{T}\alpha \nonumber\\ 
&= \alpha^{T} \mathcal{E} \alpha - \alpha^{T}\mathcal{E}_{K}P^{-1}\mathcal{E}_{K}^{T}\alpha\nonumber\\
&\leq \alpha^{T} \mathcal{E} \alpha \hspace{25pt}     \text{          (Since, P is an SPD matrix, $\alpha^{T}\mathcal{E}_{K}P^{-1}\mathcal{E}_{K}^{T}\alpha > 0$)} \label{compactError}\\
&\text{We seek for a bound for the above expression. For that we observe the following:} \nonumber\\
&(\mathcal{E})_{ik} = \langle f_{i}(t-t_i)-K^{j_i}(t-t_{i}^{'}),f_{k}(t-t_k)-K^{j_k}(t-t_{k}^{'})\rangle \nonumber\\
&=||f_{i}(t-t_i)-K^{j_i}(t-t_{i}^{'})||_{2}.||f_{k}(t-t_k)-K^{j_k}(t-t_{k}^{'})||_{2}.x_{ik} \nonumber\\
& \text{(where $x_{ik} \in [0,1]$. We also note that $x_{ik}$ is close to $0$ when there is not much overlap in the} \nonumber\\ 
& \text{support of the two components and their corresponding fitting kernels.)} \nonumber\\
&=x_{ik}.(||(f_{i}(t-t_i)-K^{j_i}(t-t_{i})||+||K^{j_i}(t-t_{i})-K^{j_i}(t-t_{i}^{'}))||) \nonumber\\ 
&\hspace{50pt}.(||f_{k}(t-t_k)-K^{j_k}(t-t_{k})||+||K^{j_k}(t-t_{k})-K^{j_k}(t-t_{k}^{'})||) \nonumber\\
&\implies (\mathcal{E})_{ik} = x_{ik}.(\delta + C\Delta)^{2} \label{eVal}\nonumber\\
\end{align}
\begin{align}
&\text{Using Gershgorin circle theorem, the maximum eigen value of $\mathcal{E}$:} \nonumber\\
&\Lambda_{max}(\mathcal{E}) \leq max_{i} ((\mathcal{E})_{ii} + \Sigma_{k \neq i}|(\mathcal{E})_{ik}|) 
\leq (\delta + C\Delta)^2 (x_{max}+1) \hspace{20pt}\text{(Using \ref{eVal})} \label{Eeig}\\
&\text{ (where $x_{max} \in [0,N-1]$ is a positive number that depends on the maximum} \nonumber\\
&\text{overlap of the supports of the component signals and their fitting kernels.)} \nonumber\\
&\text{Similarly, the minimum eigen value of $F$ is:}
\nonumber\\
&\Lambda_{min}(F) =  min_{i} ((F)_{ii} - \Sigma_{i \neq k} |\langle f_{p_i}(t-t_i), f_{p_k}(t-t_k)\rangle|) \leq 1- \eta \label{Feig}\\
& \text{(Since by assumption $\Sigma_{i \neq k} |<f_{p_i}(t-t_i), f_{p_k}(t-t_k)>|  \leq \eta$ )} \nonumber\\
&\text{Combining the results from \ref{compactError}, \ref{Eeig} and \ref{Feig} we get:} \nonumber\\
&||X(t)- X_{hyp}(t)||_{2}^{2}/||X(t)||_{2}^{2} \leq \alpha^{T} \mathcal{E} \alpha / \alpha^{T} F \alpha 
\leq \Lambda_{max}(\mathcal{E})/\Lambda_{min}(F) \nonumber\\
&\leq  (\delta + C\Delta)^2 (x_{max}+1) /(1- \eta)\\
&\text{Finally using \ref{part1} we conclude,} \nonumber\\
&||X(t)- X^{*}(t)||_{2}^{2}/||X(t)||_{2}^{2} \leq
||X(t)- X_{hyp}(t)||_{2}^{2}/||X(t)||_{2}^{2}
\leq
(\delta + C\Delta)^2 (x_{max}+1) /(1- \eta) \nonumber\\\nonumber
\end{align}
\section{Kernel Adaptation}
\label{Learning}

As demonstrated by the perfect reconstruction theorem, over a given class of input signals ${F}$, reconstructions are perfect when the kernels from the ensemble $K$ match the (unknown) underlying components that generate the signals in ${F}$. With the goal of improving the quality of the reconstruction, we now propose a gradient descent based method that incrementally changes the kernels to decrease reconstruction error, which then indirectly induces the kernels to fit the unknown components from which signals in ${F}$ are generated. 

Informally speaking, there are two effects that a perturbation of a kernel has on the reconstructed signal: (i) perturbation of the kernel $K^{j_i}$ directly impacts the reconstruction $\sum \alpha_i K^{j_i}$, and (ii) perturbation of the kernel perturbs the spike times which then have an impact on the locations at which the kernels are situated to be summed. The perturbation of the spike times also incurs a ``domino effect'' on future spike times via their ahps. 

To be able to apply gradient descent on the kernels, we consider the kernels in their parametric forms, i.e., the kernel response function of $K^{j}$ is assumed to be of the form, $K^j(t) = K^j(t;\{\theta^{j}_c\})$, where $\{\theta^{j}_c\}$ are the free parameters that govern the shape of the kernel $K^{j}$. In this context, our goal is to derive an update rule for the free parameters $\theta^{j}_c$s.

Formally, let us consider a family of signals, $F$, and let $X^*(t)$ be the reconstruction of $X(t)$ for $X(t) \in F$. The goal is to arrive at the optimal set of kernels that minimize the expected reconstruction error $\int ||X(t) - X^*(t)||^2 dF_X $ using gradient descent, where $F_X$ is the cumulative probability distribution over $F$ at  $X(t)$. Shifting to a stochastic gradient descent framework, for a randomly selected signal $X(t) \in F$, the reconstruction error of $X(t)$, using (\ref{eq:reconsSignalEq}), is defined as:
\begin{equation}
\label{eq:errorDef}
	E_X =  ||X(t) - X^*(t)||^2 = \int (X^*(\tau) - X(\tau))^2 d\tau
 = \int (\sum\limits_{i=1}^{N} [ \alpha_i k^{j_i}(t_i-\tau)] - X(\tau)) ^2 d\tau
\end{equation}
\paragraph{Derivative of $t_i$.} First, we notice that each spike is generated by a certain neuron, and only the $\theta^j_c$s of the kernel response of that neuron have an effect on the time of the spike. In other words, if $j_i \ne k$ then $\forall{c} ; \nicefrac{\partial t_i}{\partial \theta_c^k} = 0$. For the other case, we can calculate $\nicefrac{\partial t_i}{\partial \theta_c^{j_i}}$ as follows: The initial condition for the generation of a spike at time $t_i$ by the $j_i$th kernel is given by equation(\ref{spikeConstraint}). Now consider that out of all the free parameters, we only perturb $\theta_c^{j_i}$ to $\theta_c^{j_i}+\Delta\theta_c^{j_i}$. This will in turn perturb all the spike times for the $j_i$th kernel. Let spike time $t_i$ be shifted to $t_i + \Delta t_i$. Let us also assume that $t_l$ is the last spike produced by $K^{j_i}$ prior to $t_i$, and $t_l$  is correspondingly shifted to $t_l+\Delta t_l$. Under this perturbed scenario, we can rewrite (\ref{spikeConstraint}) as:  
\begin{equation}
\label{updated}
	\int X(\tau) K^{j_i}(t_i + \Delta t_i -\tau;\theta_c^{j_i}+\Delta\theta_c^{j_i}) d\tau = (T+ \Delta T)^{j_i}(t_i+\Delta t_i) 
\end{equation}
Combining (\ref{spikeConstraint}) and (\ref{updated}), using  Taylor series approximation, ignoring all second and higher order terms, using the form of the simple threshold function $T^{j_i}(t)$ as in (\ref{thresholdeq}) and setting $\lim{\Delta\theta_c^{j_i}\to 0}$, the derivative of $t_i$ with respect to a particular kernel parameter $\theta_c^{j_i}$ is obtained as:
\begin{equation}
\label{timeDerivative}
    \frac{\partial t_i}{\partial \theta_c^{j_i}} = 
    \begin{cases*}
        - \frac{ \int X(\tau)  \frac{\partial K^j_i(t;\theta_c^{j_i})}{\partial \theta_c^{j_i}}\Big|_{t=t_i-\tau} d\tau - 
     \frac{M^{j_i}- C^{j_i}}{\delta_{j_i}}\frac{\partial t_k}{\partial \theta_c^{j_i}}}
{\int(\frac{\partial X(t)}{\partial t}\Big|_{t=\tau}K^j_i(t_i-\tau)d\tau +
\frac{M^{j_i}- C^{j_i}}{\delta_{j_i}}} , &  $t_i-\delta_{j} \leq t_l$ \\
 - \frac{ \int X(\tau)  \frac{\partial K^j_i(t;\theta_c^{j_i})}{\partial \theta_c^{j_i}}\Big|_{t=t_i-\tau} d\tau}
{\int(\frac{\partial X(t)}{\partial t}\Big|_{t=\tau}K^j_i(t_i-\tau)d\tau} , &  $t_i-\delta_{j} > t_l$ \\
    \end{cases*}
\end{equation}
It is important to note that the formula for $\frac{\partial t_i}{\partial \theta_c^{j_i}}$ involves derivative of previous spike times, i.e., $\frac{\partial t_l}{\partial \theta_c^{j_i}}$s and hence can be computed recursively. This is the ``domino effect'' referred to earlier. 
\paragraph{Derivative of $E_X$.} Using equation (\ref{eq:errorDef}), we can find the derivative of the error functional with respect to the free parameters as:
\begin{equation*}
	\frac{\partial E_X}{\partial \theta_c^j} =  
	\sum_{i=1}^{N} 2\frac{\partial \alpha_i}{\partial \theta_c^j} \int [X^*(\tau) - X(\tau)] K^{j_i}(t_i-\tau) d\tau +
\end{equation*}
	\begin{multline}
	\label{finalError}
		\sum_{i=1}^{N} 2\alpha_i\int [X^*(\tau) - X(\tau)] \frac{\partial K^{j_i}(t)}{\partial t}\Big|_{t=t_i-\tau}\frac{\partial t_i}{\partial \theta_c^{j}} d\tau +\\
		\sum_{i=1}^{N} 2\alpha_i\int [X^*(\tau) - X(\tau)] \frac{\partial K^{j_i}(t;\{\theta_c^{j_i}\})}{\partial \theta_c^{j}}\Big|_{t=t_i-\tau}d\tau
\end{multline}
Using the constraint in equation (\ref{optimizationproblem}), we see that the first term above goes to $0$, leaving the other two remaining terms. We emphasize that this corresponds to the fact that $\frac{\partial E_X}{\partial \theta_c^j}$ depends upon the $\alpha_i$s but not on the $\frac{\partial \alpha_i}{\partial \theta_c^j}$.

For a chosen set of kernels, given in a particular parametric form, equations (\ref{finalError}) and (\ref{timeDerivative}) hold all the information necessary to apply stochastic gradient descent on the $\theta_c^{j}$s to minimize the expected reconstruction error.

\section{Experiments}
\label{Experiments}

The experiments were targeted toward establishing the effectiveness of the proposed learning technique. To achieve this, we first chose an ensemble of $n$ kernels $K$ in a specific parametric form. Here, we used B-splines of order 3 to construct our kernels. Specifically, for kernel $K^{j} \in K$, its response function was of the form:\\
$$
K^j(t) = \sum_{c=1}^{C}\beta^j_cB(\alpha^j(t-\delta^j_c))\hspace{10pt}(C \in Z^{+}) \nonumber
$$
where, $B(t)$ is the standard continuously differentiable B-spline function of order 3. In our experiments $\beta^j_c$s $\in R$ were chosen as the free parameters for the kernels and were randomly initialized. The rest of the parameters were kept fixed throughout the experiments. Next, to construct a class of input signals ${F}$ another ensemble of $n$ functions ${J}$ built out of similar B-splines but with different values of the free parameters was chosen. Specifically an $f^{j}(t) \in {J}$ was given by:
$$
f^j(t) = \sum_{c=1}^{C}\gamma^j_cB(\alpha^j(t-\delta^j_c)) \hspace{10pt}(C \in Z^{+}, \gamma^j_c  \in R \text{  and in general } \gamma^j_c \ne \beta^j_c)\\
$$
Using this ensemble of functions $J$, the class of input signals ${F}$ were constructed as
a linear combinations of randomly shifted $f^{j}(t)$s, i.e., any $X(t) \in {F}$ was of the form:\\
$$
   {F}= \{X(t)| X(t) = \sum_{i=1}^{N} a_i f^{j_i}(t_i-t) ,(\text{ for some } N \in Z^{+}, a_i \in R, t_i \in R, f^{j_i} \in J\} 
$$
 In the experiments, at each step a random sample $X(t)$ from the class ${F}$ was chosen. We then made a gradient update to the free parameters $\beta^j_c$s of the kernels $K^{j}$s as described in Section~\ref{Learning}. If the proposed learning were effective, the reconstruction error would drop rapidly resulting in near perfect reconstruction. We found this to be the case in all our experiments, as reported below.

\begin{figure}[h]
 	\centering
 	\subfloat[]{{\includegraphics[width=42mm]{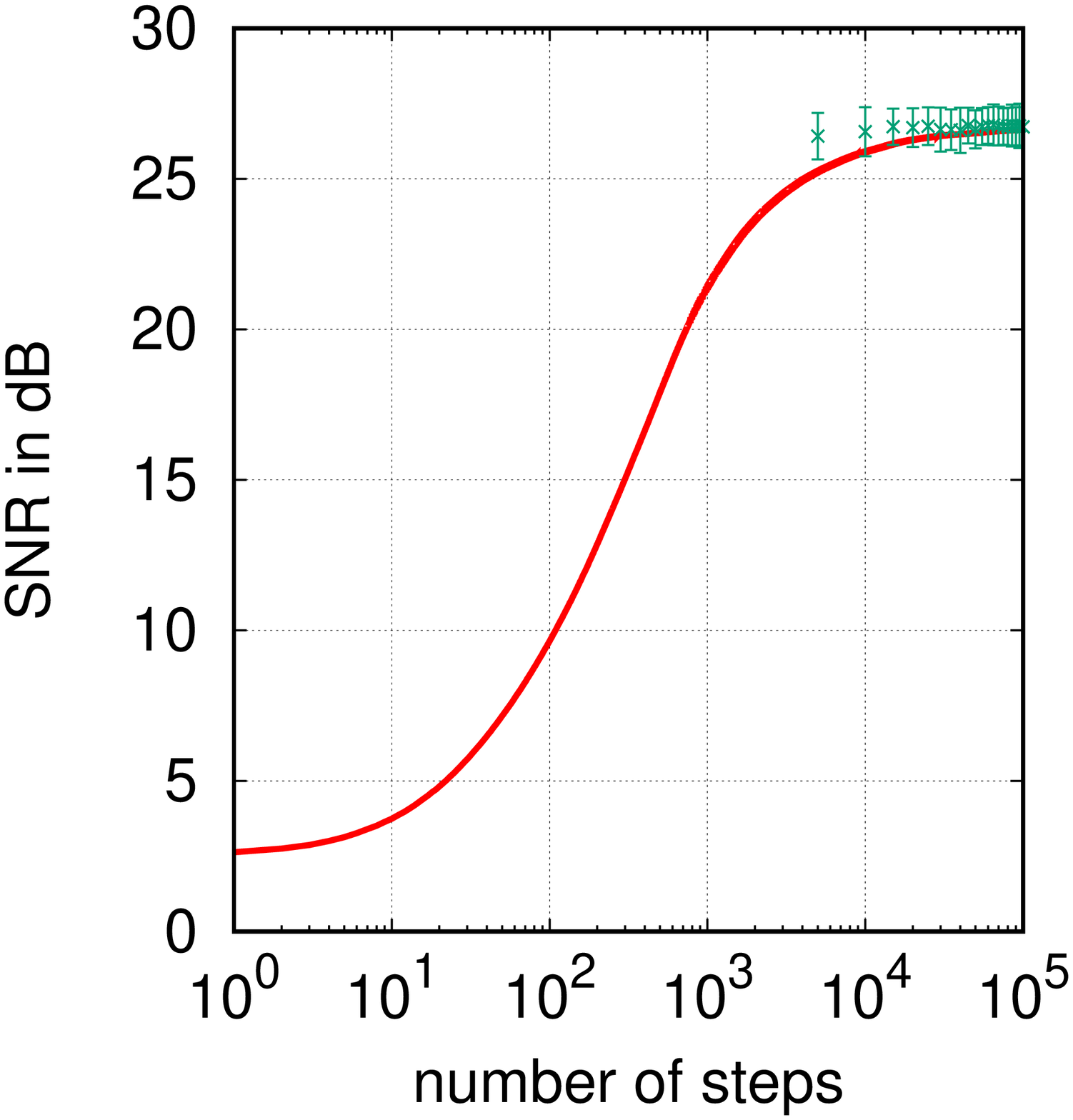} }}%
 	\quad
 	\subfloat[]{{\includegraphics[width=42mm]{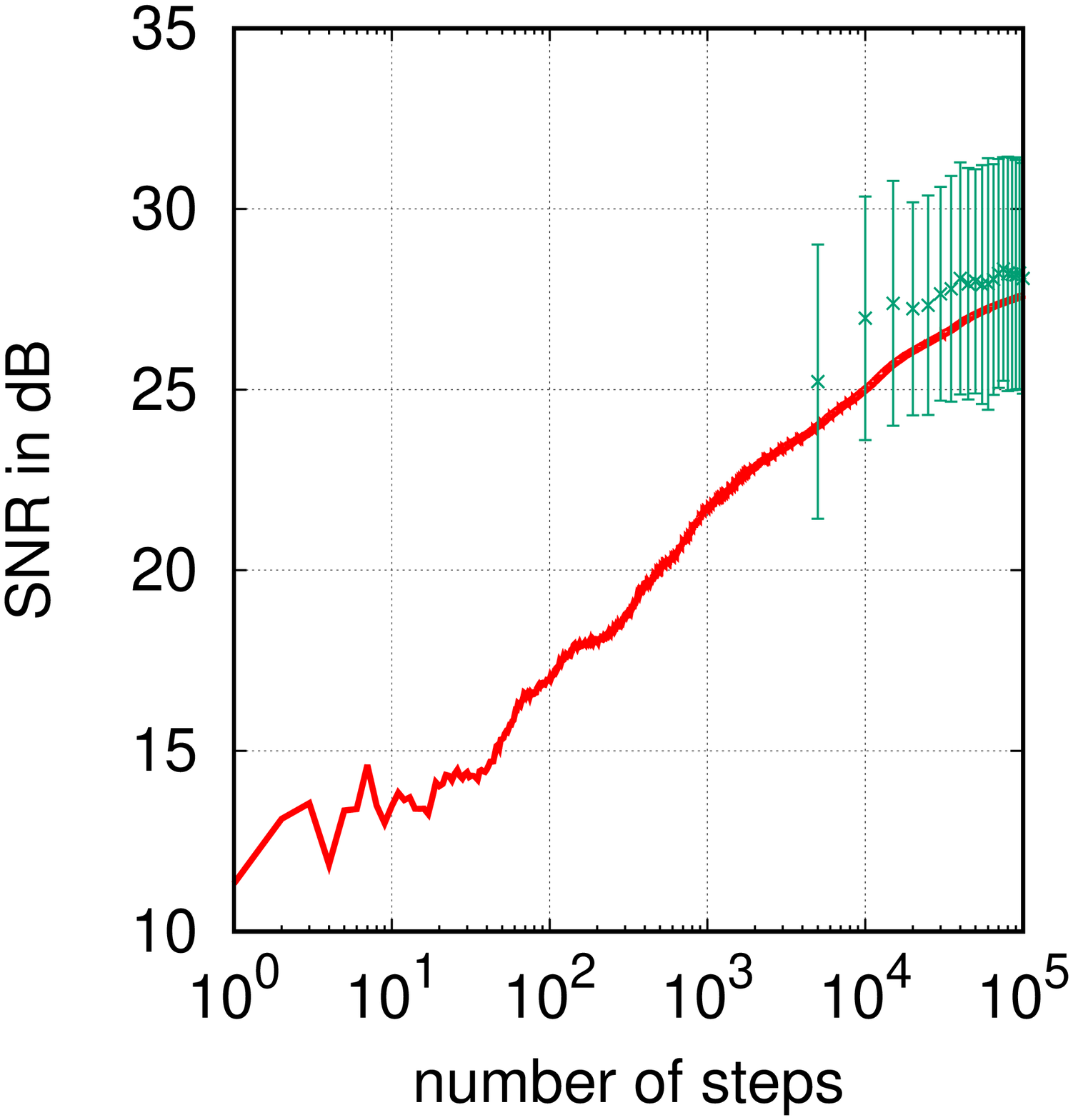} }}%
 	\quad
 	\subfloat[]{{\includegraphics[width=42mm]{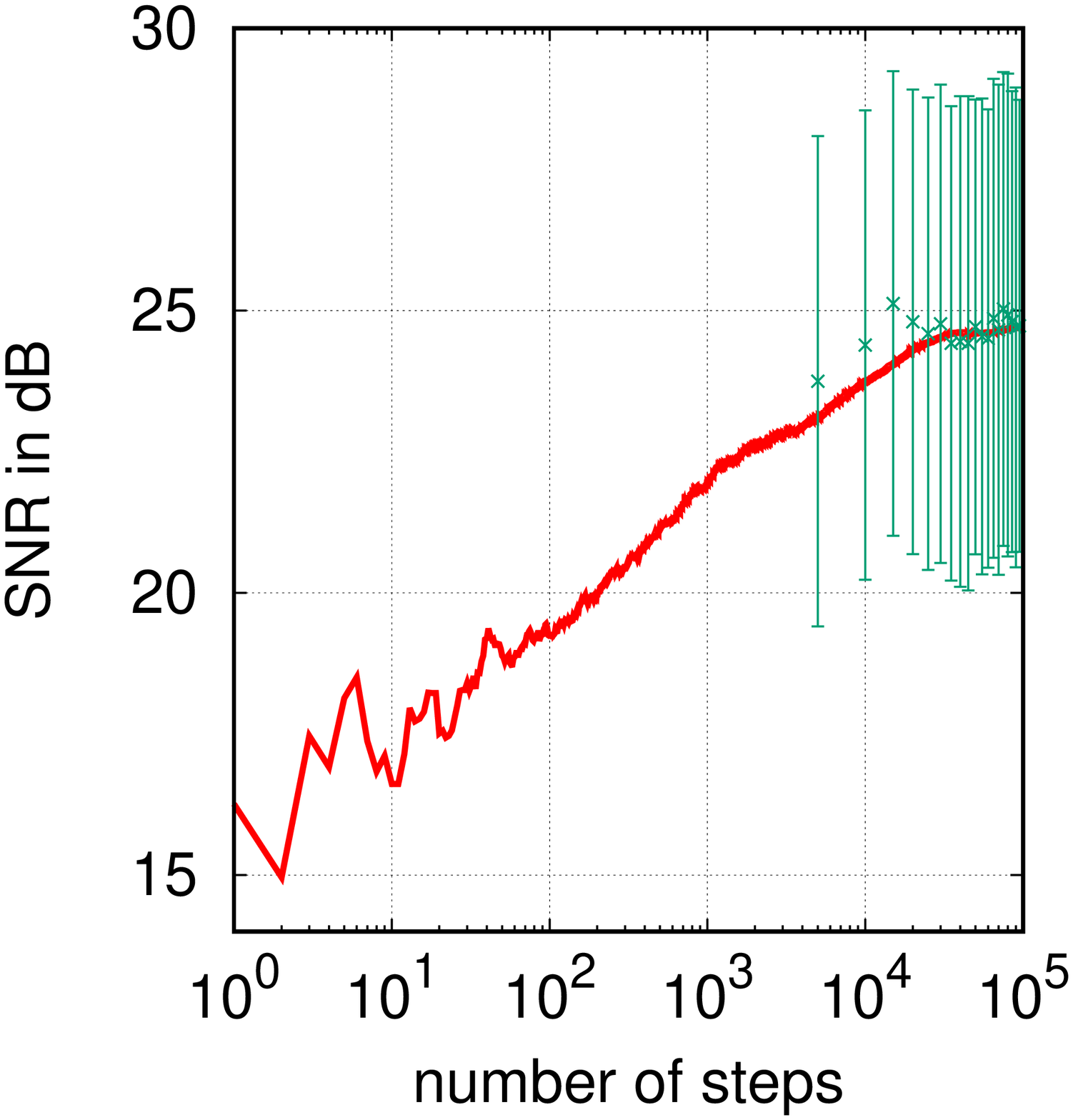} }}%
 	\caption{shows the moving averages of SNR values (in dB scale) of reconstruction errors during the training of the kernels for (a) 1 kernel (b) 5 kernels and (c) 10 kernels, plotted against the number of iterations (shown in log-scale). The error-bars (in green) show the mean and standard deviation of SNR values in reconstructions during testing after completion of a number of learning steps given by the corresponding value on the x-axis.}%
 	\label{LearningErrorCurve}
\end{figure}

In repeated experiments, and across varied number ($n$) of differently initialized kernels, we saw steady drop in reconstruction error and with sufficiently large number of learning iterations reconstruction became perfect. Here, we have reported results of sample experiments with 1,5 and 10 kernels in Figure ~\ref{LearningErrorCurve}. In each case at least $100K$ iterations of learning were executed and after each $1000$ steps of training, kernels were extracted to be run against $1000$ randomly generated new input samples for reporting testing error. The mean reconstruction error on the test set and its standard deviation have been reported at regular intervals using error-bars in the same Figure ~\ref{LearningErrorCurve}.  

One of the primary reasons behind choosing B-splines was its universal approximation property in that any arbitrary bounded-time continuous function can be approximated by B-splines. In that sense a demonstration of learning with B-spline kernels establishes generality. For learning to work we had to tune the parameters properly (e.g. setting \emph{ahp} refractory period, initial thresholds to low values, putting a reasonable learning rate, etc.) all of which can be found in configurations in our java-based implementation. All experiments were run on standard 8GB machines with quad-core processors taking reasonable time($\sim 1$sec per iteration).
\section{Conclusion}
\label{Conclusion}

In this paper, we have formulated a framework that identifies the precise conditions under which continuous time signals can be represented using an ensemble of spike trains, from which the signal can be recovered perfectly. Although aligned in their goals, this framework is very different from that investigated in Niquist-Shanon theory. The primary difference between the two lies in their respective modes of representation/coding. Instead of sampling the value of a function at uniform or non-uniform prespecified sample points the new coding scheme reports the (non-uniform) sample points where the function takes specific convolved values. A natural extension of this framework addresses approximate reconstruction, both when there is noise injected in the spike timings as well as in the kernel functions. Coding is intimately related to compression and our experimental results indicate great potential in this regard. The simulation source code is available at: \url{bitbucket.org/crystalonix/oldsensorycoding.git}.
\small
{\small\bibliography{paper.bib}}

\begin{thebibliography}{20}
\providecommand{\natexlab}[1]{#1}
\providecommand{\url}[1]{\texttt{#1}}
\expandafter\ifx\csname urlstyle\endcsname\relax
  \providecommand{\doi}[1]{doi: #1}\else
  \providecommand{\doi}{doi: \begingroup \urlstyle{rm}\Url}\fi

\bibitem[Barlow(1961)]{barlow1961}
Horace~B Barlow.
\newblock Possible principles underlying the transformations of sensory
  messages.
\newblock \emph{Sensory Communication}, pages 217--234, 1961.

\bibitem[Boyd and Chua(1985)]{boyd1985}
Stephen Boyd and Leon Chua.
\newblock Fading memory and the problem of approximating nonlinear operators
  with volterra series.
\newblock \emph{IEEE Transactions on circuits and systems}, 32\penalty0
  (11):\penalty0 1150--1161, 1985.

\bibitem[C~Smith and Lewicki(2006)]{CSmith2006}
Evan C~Smith and Michael Lewicki.
\newblock Efficient auditory coding.
\newblock \emph{Nature}, 439:\penalty0 978--82, Mar 2006.

\bibitem[Chklovskii and Soudry(2012)]{chklovskii}
Dmitri~B. Chklovskii and Daniel Soudry.
\newblock Neuronal spike generation mechanism as an oversampling, noise-shaping
  a-to-d converter.
\newblock In F.~Pereira, C.~J.~C. Burges, L.~Bottou, and K.~Q. Weinberger,
  editors, \emph{Advances in Neural Information Processing Systems 25}, pages
  503--511. 2012.

\bibitem[Christopher~deCharms and Merzenich(1996)]{christopher1996}
R~Christopher~deCharms and Michael~M Merzenich.
\newblock Primary cortical representation of sounds by the coordination of
  action-potential timing.
\newblock \emph{Nature}, 381\penalty0 (6583):\penalty0 610, 1996.

\bibitem[Daubechies and DeVore(2003)]{daubechies}
Ingrid Daubechies and Ron DeVore.
\newblock Approximating a bandlimited function using very coarsely quantized
  data: A family of stable sigma-delta modulators of arbitrary order.
\newblock \emph{Annals of Mathematics}, 158\penalty0 (2):\penalty0 679--710,
  2003.

\bibitem[F{\"o}ldi{\'a}k(1990)]{Foldiak1990}
P.~F{\"o}ldi{\'a}k.
\newblock Forming sparse representations by local anti-hebbian learning.
\newblock \emph{Biological Cybernetics}, 64\penalty0 (2):\penalty0 165--170,
  Dec 1990.

\bibitem[Graham and Field(2007)]{Graham}
Daniel Graham and David Field.
\newblock Sparse coding in the neocortex.
\newblock \emph{Evolution of Nervous Systems}, 3:\penalty0 181--187, 2007.

\bibitem[Johansson and Birznieks(2004)]{johansson2004}
Roland~S Johansson and Ingvars Birznieks.
\newblock First spikes in ensembles of human tactile afferents code complex
  spatial fingertip events.
\newblock \emph{Nature neuroscience}, 7\penalty0 (2):\penalty0 170--177, 2004.

\bibitem[Lewicki(2002)]{Lewicki2002}
Michael~S. Lewicki.
\newblock Efficient coding of natural sounds.
\newblock \emph{Nature Neuroscience}, 5:\penalty0 356--363, Mar 2002.

\bibitem[Mallat and Zhang(1993)]{Mallat1993}
S.~G. Mallat and Zhifeng Zhang.
\newblock Matching pursuits with time-frequency dictionaries.
\newblock \emph{IEEE Transactions on Signal Processing}, 41\penalty0
  (12):\penalty0 3397--3415, 1993.

\bibitem[Meister et~al.(1995)Meister, Lagnado, Baylor, et~al.]{meister1995}
Markus Meister, Leon Lagnado, Denis~A Baylor, et~al.
\newblock Concerted signaling by retinal ganglion cells.
\newblock \emph{Science}, 270\penalty0 (5239):\penalty0 1207--1210, 1995.

\bibitem[Nemenman et~al.(2008)Nemenman, Lewen, Bialek, and van
  Steveninck]{nemenman2008}
Ilya Nemenman, Geoffrey~D Lewen, William Bialek, and Rob R de~Ruyter van
  Steveninck.
\newblock Neural coding of natural stimuli: information at sub-millisecond
  resolution.
\newblock \emph{PLoS computational biology}, 4\penalty0 (3):\penalty0 e1000025,
  2008.

\bibitem[Neuenschwander and Singer(1996)]{neuenschwander1996}
Sergio Neuenschwander and Wolf Singer.
\newblock Long-range synchronization of oscillatory light responses in the cat
  retina and lateral geniculate nucleus.
\newblock \emph{Nature}, 379\penalty0 (6567):\penalty0 728, 1996.

\bibitem[Olshausen(1996)]{Olshausen1996}
David~J. Olshausen, Bruno A.and~Field.
\newblock Emergence of simple-cell receptive field properties by learning a
  sparse code for natural images.
\newblock \emph{Nature}, 381:\penalty0 607--609, Jun 1996.

\bibitem[Sch{\"o}lkopf et~al.(2001)Sch{\"o}lkopf, Herbrich, and
  Smola]{scholkopf2001}
Bernhard Sch{\"o}lkopf, Ralf Herbrich, and Alex~J Smola.
\newblock A generalized representer theorem.
\newblock In \emph{International Conference on Computational Learning Theory},
  pages 416--426. Springer, 2001.

\bibitem[Shapero et~al.(2014)Shapero, Zhu, Hasler, and Rozell]{shapero}
Samuel Shapero, Mengchen Zhu, Jennifer Hasler, and Christopher Rozell.
\newblock Optimal sparse approximation with integrate and fire neurons.
\newblock \emph{International journal of neural systems}, 24:\penalty0 1440001,
  08 2014.

\bibitem[Squire(2008)]{squire}
L.R. Squire.
\newblock \emph{Fundamental Neuroscience}.
\newblock Academic Press/Elsevier, 2008.
\newblock ISBN 9780123740199.

\bibitem[Wehr and Laurent(1996)]{wehr1996}
Michael Wehr and Gilles Laurent.
\newblock Odour encoding by temporal sequences of firing in oscillating neural
  assemblies.
\newblock \emph{Nature}, 384\penalty0 (6605):\penalty0 162, 1996.

\bibitem[Zylberberg et~al.(2011)Zylberberg, Murphy, and DeWeese]{murphy}
Joel Zylberberg, Jason~Timothy Murphy, and Michael~Robert DeWeese.
\newblock A sparse coding model with synaptically local plasticity and spiking
  neurons can account for the diverse shapes of v1 simple cell receptive
  fields.
\newblock \emph{PLOS Computational Biology}, 7\penalty0 (10):\penalty0 1--12,
  10 2011.

\end{thebibliography}
\bibliographystyle{plainnat}





\end{document}